\documentclass[aps,prl,twocolumn,superscriptaddress,floatfix,longbibliography]{revtex4-2}

\usepackage[letterpaper, total={6in, 8in}]{geometry}
\usepackage{float}
\usepackage{amsmath}%
\usepackage{amsfonts}%
\usepackage{amssymb}%
\usepackage{amstext}
\usepackage{textgreek}
\usepackage{graphicx}
\usepackage[table,xcdraw]{xcolor}
\usepackage[colorlinks=true,allcolors=blue]{hyperref}
\usepackage{color}
\usepackage{orcidlink}
\usepackage{xcolor}
\usepackage{booktabs}
\usepackage[page,toc,titletoc,title]{appendix}
\makeatletter
\renewcommand{\p@subsection}{}
\renewcommand{\p@subsubsection}{}
\makeatother

\usepackage{xspace} %CPM: there is debate about if xpspace causes more problems than it solves.
\usepackage{pstricks,pstricks-add} 
\begin{document}
\title{Compressional heating above 1 keV on the LM26 magnetized target fusion machine}

\author{ S.~J. Howard~\orcidlink{0009-0007-3247-6096}}
\affiliation{General Fusion Inc., Richmond, British Columbia, Canada}
\author{D.~Krotez~\orcidlink{0009-0000-8302-6879}}
\affiliation{General Fusion Inc., Richmond, British Columbia, Canada}
\author{ P.~Carle~\orcidlink{0000-0003-4317-1451}}
\affiliation{General Fusion Inc., Richmond, British Columbia, Canada}
\author{J.~Sanchez Rojo}
\affiliation{General Fusion Inc., Richmond, British Columbia, Canada}
\author{R.~Underwood}
\affiliation{General Fusion Inc., Richmond, British Columbia, Canada}
\author{ A.~Froese~\orcidlink{0000-0002-1669-7397}}
\affiliation{General Fusion Inc., Richmond, British Columbia, Canada}
\author{ M.~Reynolds~\orcidlink{0000-0001-5880-2290}}
\affiliation{General Fusion Inc., Richmond, British Columbia, Canada}
\author{ N.~Sirmas }
\affiliation{General Fusion Inc., Richmond, British Columbia, Canada}
\author{K.~Conquergood}
\affiliation{General Fusion Inc., Richmond, British Columbia, Canada}
\author{ K.~Epp }
\affiliation{General Fusion Inc., Richmond, British Columbia, Canada}
\author{ P.~Forysinski }
\affiliation{General Fusion Inc., Richmond, British Columbia, Canada}
\author{ D.~Plant }
\affiliation{General Fusion Inc., Richmond, British Columbia, Canada}
\author{ R.~Zindler }
\affiliation{General Fusion Inc., Richmond, British Columbia, Canada}
\author{ E.~Love }
\affiliation{General Fusion Inc., Richmond, British Columbia, Canada}
\author{ D.~P.~Brennan }
\affiliation{General Fusion Inc., Richmond, British Columbia, Canada}
\author{ N.~Kumar~\orcidlink{0000-0001-8693-555X}}
\affiliation{General Fusion Inc., Richmond, British Columbia, Canada}
\author{C.~MacDonald }
\affiliation{General Fusion Inc., Richmond, British Columbia, Canada}
\author{ A.~Wong}
\affiliation{General Fusion Inc., Richmond, British Columbia, Canada}
\author{ W.~Zawalski}
\affiliation{General Fusion Inc., Richmond, British Columbia, Canada}
\author{ B.~Rablah}
\affiliation{General Fusion Inc., Richmond, British Columbia, Canada}
\author{ W.~Kozicki}
\affiliation{General Fusion Inc., Richmond, British Columbia, Canada}
\author{ C.~Preston}
\affiliation{General Fusion Inc., Richmond, British Columbia, Canada}
\author{A.~Lee}
\affiliation{General Fusion Inc., Richmond, British Columbia, Canada}
\author{X.~Feng}
\affiliation{General Fusion Inc., Richmond, British Columbia, Canada}
\author{M. Schellenberg-Beaver}
\affiliation{General Fusion Inc., Richmond, British Columbia, Canada}
\author{R.~Tingley}
\affiliation{General Fusion Inc., Richmond, British Columbia, Canada}
\author{ J.~Gorenstein}
\affiliation{General Fusion Inc., Richmond, British Columbia, Canada}
\author{ V.~Suponitsky }
\affiliation{General Fusion Inc., Richmond, British Columbia, Canada}
\author{Z.~Seifollahi Moghadam }
\affiliation{General Fusion Inc., Richmond, British Columbia, Canada}
\author{ D.~Froese}
\affiliation{General Fusion Inc., Richmond, British Columbia, Canada}
\author{E.~Cessford }
\affiliation{General Fusion Inc., Richmond, British Columbia, Canada}
\
\author{J.~Pratt}
\affiliation{General Fusion Inc., Richmond, British Columbia, Canada}

\author{ J.~Crofts}
\affiliation{General Fusion Inc., Richmond, British Columbia, Canada}
\author{ J.~Sardari}
\affiliation{General Fusion Inc., Richmond, British Columbia, Canada}

\author{G.~Faust}
\affiliation{General Fusion Inc., Richmond, British Columbia, Canada}
\author{D.~Ross}
\affiliation{General Fusion Inc., Richmond, British Columbia, Canada}
\author{ J.~Wilkie}
\affiliation{General Fusion Inc., Richmond, British Columbia, Canada}
\author{S.~Bernard}
\affiliation{General Fusion Inc., Richmond, British Columbia, Canada}
\author{ C.~Connor}
\affiliation{General Fusion Inc., Richmond, British Columbia, Canada}
\author{S.~Bolanos }
\affiliation{General Fusion Inc., Richmond, British Columbia, Canada}

\author{ C.~Gutjahr}
\affiliation{General Fusion Inc., Richmond, British Columbia, Canada}

\author{ E.~Chan}
\affiliation{General Fusion Inc., Richmond, British Columbia, Canada}

\author{ C.~Eyrich }
\affiliation{General Fusion Inc., Richmond, British Columbia, Canada}

\author{ A.~Mahoney }
\affiliation{General Fusion Inc., Richmond, British Columbia, Canada}

\author{M.~Davidson}
\affiliation{General Fusion Inc., Richmond, British Columbia, Canada}

\author{ M.~Laberge~\orcidlink{0009-0006-0093-2168}} 
\affiliation{General Fusion Inc., Richmond, British Columbia, Canada}

\begin{abstract}

The LM26 device has achieved a peak electron temperature of $T_e = 1180 \pm 65$\,eV as measured by filtered X-ray diodes, supported by adjacent Thomson scattering measurements of $T_e = 1090 \pm 40$\,eV, with a deuterium ion temperature of $T_i = 459 \pm 33$\,eV as inferred from neutron yield.  LM26 compresses a spherical tokamak plasma inside an initially 1.76 m diameter solid lithium liner imploded by theta-pinch coils.  Radial compression by a factor of 2.75 increases $T_e$ over fivefold, while $T_i$ increases as much as twofold. The integrated observation of magnetic flux compression, electron heating, ion heating, and neutron production in a magnetized plasma compressed by a large lithium liner is an important milestone for magnetized target fusion.

\end{abstract}
\date{\today}

\maketitle

The Lawson Machine 26 (LM26) plasma compression experiment operating at General Fusion has demonstrated heating a plasma to fusion-relevant temperatures greater than 1\,keV using compression by a lithium cylindrical shell (liner)
as the primary means of heating.
This machine has successfully applied 12.6\,MW of compressional heating power to the plasma, demonstrating key aspects of the magnetized target fusion (MTF) concept as a viable path to commercial fusion energy.
The imploding liner serves multiple purposes simultaneously, providing a cost-effective heating source for a pulsed fusion burn while also acting as a passive magnetic flux conserver and coaxial current return path for a symmetric single-turn toroidal field coil \cite{Science_of_LM26_Howard_NucFus_2026}.  
In a power plant, a collapsing liquid lithium liner serves the same purpose, and would additionally serve as a neutron-absorbing first wall and breeding blanket that radially thickens as the compression proceeds.  
This holistic approach addresses many of the most challenging practical issues that are anticipated with more standard approaches of tokamaks and stellarators, such as first wall survivability and neutron damage to reactor structures, which will become increasingly severe as operation is pushed into the burning plasma regime necessary for commercial operation \cite{rowcliffe_materials_2018, linke_challenges_2019, zinkle_fusion_2005}.

The initial results from LM26 experimental operations have been presented in detail in Howard et al.\ \cite{Science_of_LM26_Howard_NucFus_2026}. LM26 is novel in its use of a solid lithium liner to compress a spherical tokamak (ST) plasma configuration and uses a non-destructive electromagnetic theta-pinch coil to drive the liner implosion. 
A Marshall gun coaxial helicity injection (CHI) system \cite{Pi3_tauE_NucFus_2025} is used to form an ST plasma into a lithium liner seated within the theta-pinch compression chamber.

This letter reports on the evidence of compressional heating of a deuterium plasma to $T_e > 1$\,keV and related ion heating and details the corresponding thermal power balance of the highest-performing compression shot so far, LMC-12 (shot 8131). The previous paper \cite{Science_of_LM26_Howard_NucFus_2026} provided detailed analysis on LMC-1 to LMC-11 with a more complete description of the experimental setup.  

\clearpage % Added this to fix inconsistent paragraph spacing on page 1

%Figure 1
\begin{figure}[tp]
    \centering
    \includegraphics[width=1\linewidth]{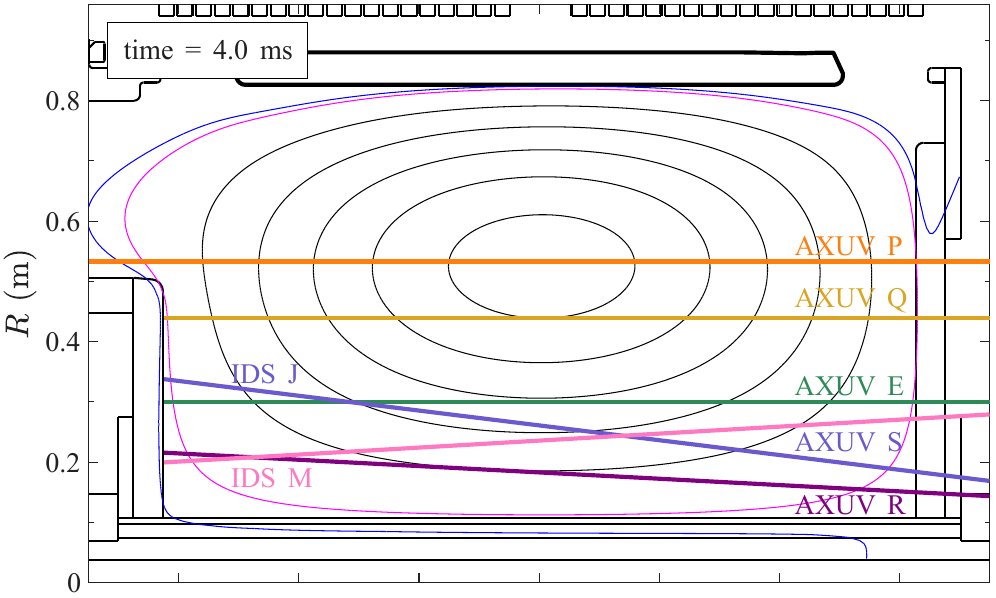}
    
    \vspace{-5pt}
    
    \includegraphics[width=1\linewidth]{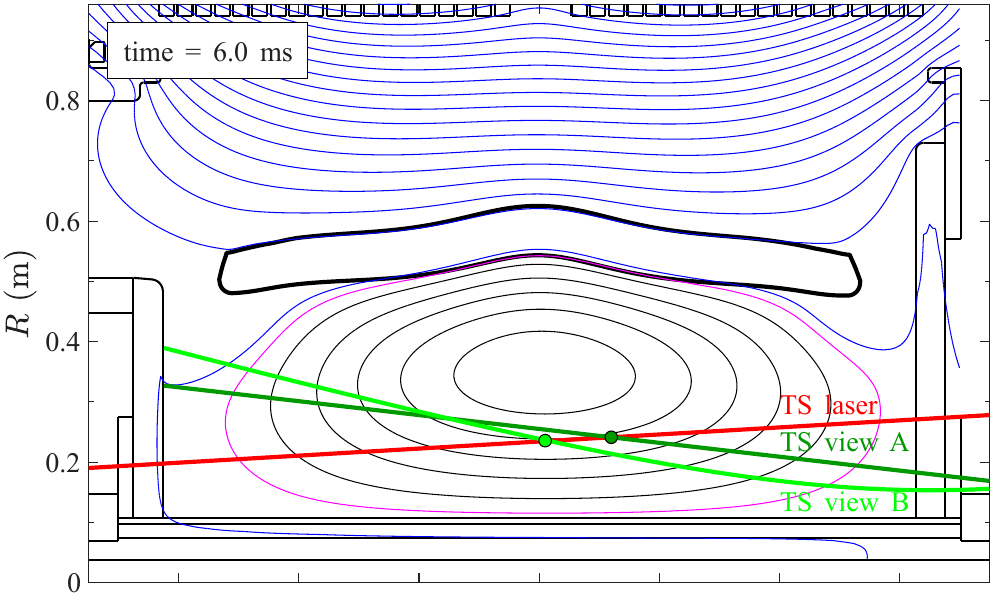}
    
    \vspace{-5pt}
    
    \includegraphics[width=1\linewidth]{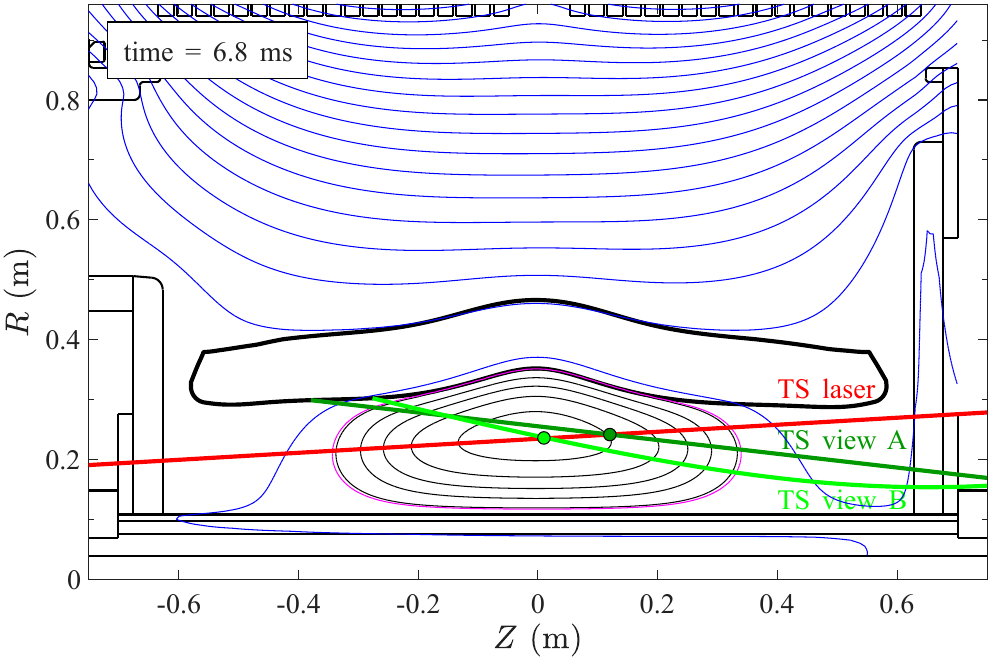}
    \caption{Evolution of plasma geometry during 
    % lithium liner compression shot 
    LMC-12 at times 4.0\,ms (compression start), 6.0\,ms (first TS pulse), and 6.8\,ms (last TS pulse). Closed poloidal flux contours are shown in black with 0.02\,Wb spacing. Open poloidal flux contours are shown in blue with 0.2\,Wb spacing. The LCFS is marked in magenta. The upper frame shows the AXUV and ion Doppler spectroscopy (IDS) chords.  The lower two frames show the TS laser and view chords with the observation points marked as circles.  Solid components are shown as black outlines.}
    \label{fig:PlasmaGeometry}
\end{figure}

%Figure 2
\begin{figure}[tp]
    \centering
    \includegraphics[width=0.99\linewidth]{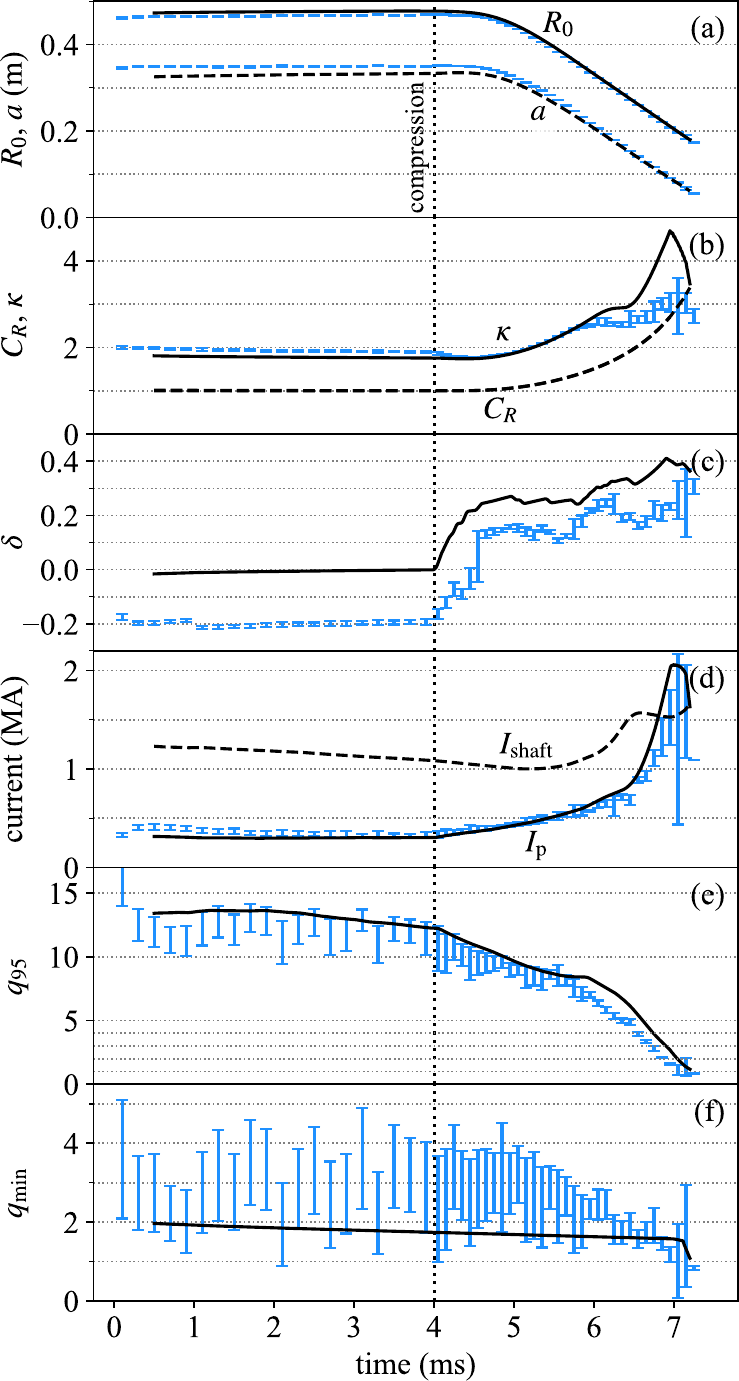}
    \caption{Plasma parameters from fits by ISM (black curves) and BPR (blue with error bars) showing (a) major and minor radii, (b) plasma elongation $\kappa$ and liner radial compression ratio $C_R$, (c) triangularity $\delta$, (d) shaft current $I_{\mathrm{shaft}}$ from toroidal Mirnov probe data and plasma current $I_{\mathrm{p}}$, (e) safety factor at the plasma edge $q_{95}$, and (f) minimum value of safety factor profile $q_{\mathrm{min}}$.  The discrepancy in $\kappa$ and $\delta$ prior to compression is due to omission of the gun opening in ISM.}
    \label{fig:PlasmaParameters}
\end{figure}

\begin{figure}
    \centering
    \includegraphics[width=1\linewidth]{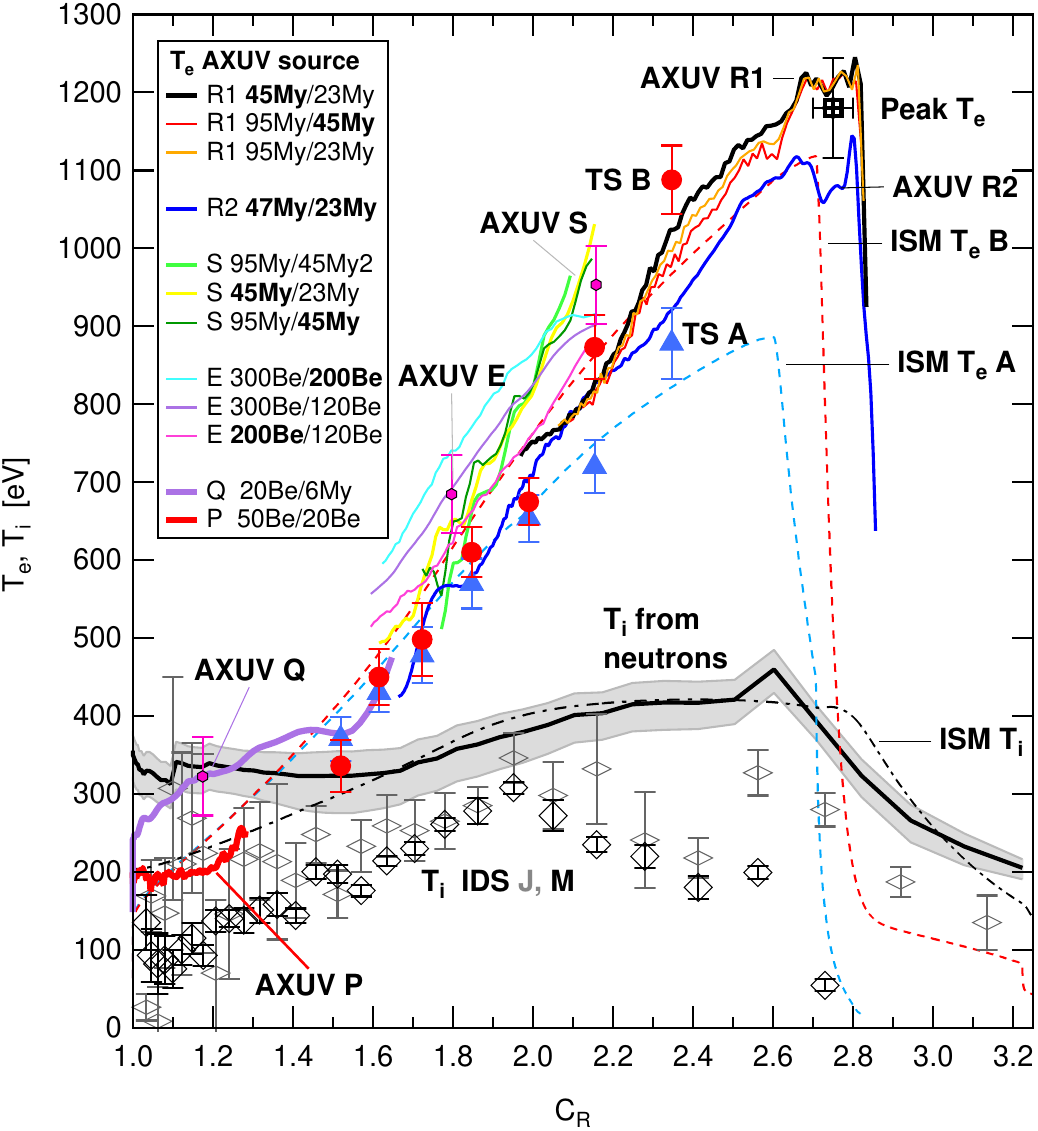}
    \caption{Summary of $T_e$ from AXUV and TS, and $T_i$ data from neutron yield and IDS, taken during the LMC-12 experiment. Temperatures are plotted as a function of $C_R$ and compared to the ISM model, which agrees well on the presence of heating to $T_e >1$\,keV. Bold AXUV legend items indicate pairs of matching filters that were averaged together as an input to $T_e$. Magenta markers show when AXUV Q, E, and S are viewing the core of the plasma. }
    \label{fig:Summarygraph}
\end{figure}

As seen in Figs.~\ref{fig:PlasmaGeometry}~and~\ref{fig:PlasmaParameters}, the inwardly accelerated lithium liner acts as a flux conserving boundary that rapidly compresses the magnetized plasma (i.e., the compression time is significantly shorter than the uncompressed plasma lifetime).  
The last closed flux surface (LCFS) is indicated on Fig.~\ref{fig:PlasmaGeometry} by the magenta contour.   
The liner radial compression ratio in Fig.~\ref{fig:PlasmaParameters}b is defined as $C_R(t) \equiv R_\mathrm{eq}(t_\mathrm{comp}) / R_\mathrm{eq}(t)$, where $R_\mathrm{eq}(t)$ is the radius of the liner at the equator and $t = t_\mathrm{comp}$ is the start of compression.
The poloidal field peaks $3.2$\,ms into compression when $C_R=3.5$, the radial midpoint of the plasma has decreased from $R_0=46.3$\,cm to 21.0\,cm, the minor radius has decreased from $a=34.9$\,cm to $9.3$\,cm, and the aspect ratio has increased from $A=1.33$ to 2.26.
Earlier when $C_R=2.75$ (3\,ms into compression) the peak temperature is attained.  After peaking, the temperature drops precipitously (Fig.~\ref{fig:Summarygraph}) as increased internal plasma resistance begins to quench the plasma current.

During the compression, plasma current $I_{\mathrm{p}}$ initially rises faster than the central shaft current $I_{\mathrm{shaft}}$. After $C_R = 1.5$, an external circuit is used to drive the shaft current above the toroidal flux conservation point based on the change in plasma-domain coaxial inductance, which has been observed to stabilize magnetohydrodynamic (MHD) instabilities that tend to occur near the mid-stage of compression \cite{Science_of_LM26_Howard_NucFus_2026,PCS16_Howard_2025}. Grad--Shafranov reconstructions indicate that the edge safety factor $q_{95}$ drops from above 10 to near 2 during compression, while the minimum of $q(\psi)$ has a more stable trend with time as shown in Fig.~\ref{fig:PlasmaParameters}.  
This is due to the core plasma following more ideal MHD behavior while the edge can resistively change $q$ value in response to $I_{\mathrm{shaft}}/I_{\mathrm{p}}$ dropping and plasma poloidal flux diffusing into the liner.   Reconstructed plasma parameters shown in Fig.~\ref{fig:PlasmaParameters} are obtained by Bayesian Plasma Reconstruction (BPR) and by applying the Integrated System Model (ISM) \cite{Science_of_LM26_Howard_NucFus_2026} as described below.

\textit{Observation of increase in $T_e$ and $T_i$---}On the LMC-12 compression we observe significant electron temperature increase from $200 \pm 22$\,eV to a peak of $1180 \pm 65$\,eV as measured by absolutely calibrated X-ray to UV sensitive diodes (AXUV), over a radial factor of 2.75 compression in 3\,ms, as shown in Fig.~\ref{fig:Summarygraph}. Plasma electron temperature measurements are made with a combination of AXUV measurements \cite{Science_of_LM26_Howard_NucFus_2026} and Thomson scattering (TS) \cite{TS_LMC12_Carle_RSI_2026}. Thomson scattered spectra are measured at 2 locations within the plasma (Fig.~\ref{fig:PlasmaGeometry}) at 8 separate times during the compression, at 125\,{\textmu}s intervals. LMC-12 was the first shot with TS view B, which better views the core late in compression and measured $T_e = 1090 \pm 40\,\mathrm{eV}$ at $C_R = 2.35$. 
The soft X-ray measurements are made on five chords using AXUV photodiode modules,
%We also have five locations of absolutely calibrated X-ray photodiode (AXUV) diagnostics, 
digitized continuously at a 250\,kHz sampling rate.
Each module has a 2$\times$2 array of filtered photodiodes, having 4 mostly-overlapping chords, each with a specific thickness of filter composed of aluminum/Mylar or beryllium. From the ratio of the signals we are able to infer the emissivity-weighted average temperature along the chord. The soft X-ray spectra were also measured for a set of 13 preparatory non-compression shots using a pulse height analysis (PHA) detector, which showed only a minimal deviation from a bremsstrahlung-like $\exp(-E_\mathrm{photon}/T_e)$ spectrum and confirmed the basis for the AXUV ratio method. 
%Figure 3

Neutron diagnostics show a $T_i$ of $323\pm39$\,eV early in compression until $C_R=1.5$, after which it increases to $459 \pm 33$\,eV at $C_R=2.6$, as shown in Fig.~\ref{fig:Summarygraph}.  ISM time-dependent reconstruction shows a twofold increase of core $T_i$
from 204\,eV to a peak of 421\,eV during compression.  
We believe the $T_i$ from neutrons is above the ISM curve before $C_R=1.5$ because a weak transient background of supra-thermal ions remaining from the fast CHI plasma formation is masking a low D-D fusion rate.

To measure the core D ion temperature in experiment, we rely on detection of D-D fusion neutrons using a spatially distributed set of eight liquid scintillators and one $^3\text{He}$ detector.
The liquid scintillators have their absolute efficiencies calculated by combining gamma calibrations of their energy response with MCNP simulations performed by the UKAEA~\cite{PCS16_Howard_2025}, which take into account neutron scattering within the moving and thickening Li liner. 
A density profile inferred from the set of interferometers allows us to infer the core $T_i$ for a given measured neutron yield using a Bosch--Hale formula \cite{Radich10022026,BoschHale1992}.
This method works for ions with a Maxwellian velocity distribution, which is plausibly satisfied for these ST plasmas with sufficient time for the ions to collisionally thermalize, given ion collision time $\tau_i\approx 0.3\mathrm{\,ms}$ at the start of compression, and the absence of auxiliary heating.
The $^3\text{He}$ detector provides a slower-responding but $\gamma$-insensitive measurement on total neutron yield using a HDPE moderator, which during the compression period ($t = 4.011$ to 7.761\,ms) measured a total yield of $(1.54 \pm 0.41)\times 10^6$\,n while PSD analysis on scintillator data during the same period gives a total yield of $(1.87 \pm 0.38)\times 10^6$\,n. 
Therefore the scintillator-based measurement of PSD neutron count rate should be accurate enough to provide $T_i(t)$ as a well-resolved function of time.  
In addition, the edge ion temperature is inferred from ion Doppler broadening of the C\,V line at 227\,nm, which also shows an increase during compression.  However, this method cannot be used to measure the core temperature due to the high $T_e$ of the core fully ionizing all carbon on the timescales of this measurement.

The suite of diagnostics also includes 17 surface-mount Mirnov probes $(B_\mathrm{pol}, B_\mathrm{tor})$ which provide the primary constraints for equilibrium reconstruction. For LMC-12 these show good core poloidal flux conservation and higher $B_\mathrm{pol}$ increase than any previous LM26 compression shot, which is consistent with electron heating resulting in decreased plasma resistivity.

\textit{Power balance analysis---}Time-dependent reconstruction is performed using ISM and the same $\chi^2$ methodology as previous LMC shots \cite{Science_of_LM26_Howard_NucFus_2026}, including the constraint of TS data. 
The plasma component of ISM is a 1.5D transport code which models the plasma as a 2D axisymmetric Grad--Shafranov equilibrium evolving via 1D flux-averaged transport processes and includes resistive diffusion of the plasma into the surrounding liner and shaft.

%Figure 4
\begin{figure}
    \centering
    \includegraphics[width=1\linewidth]
    {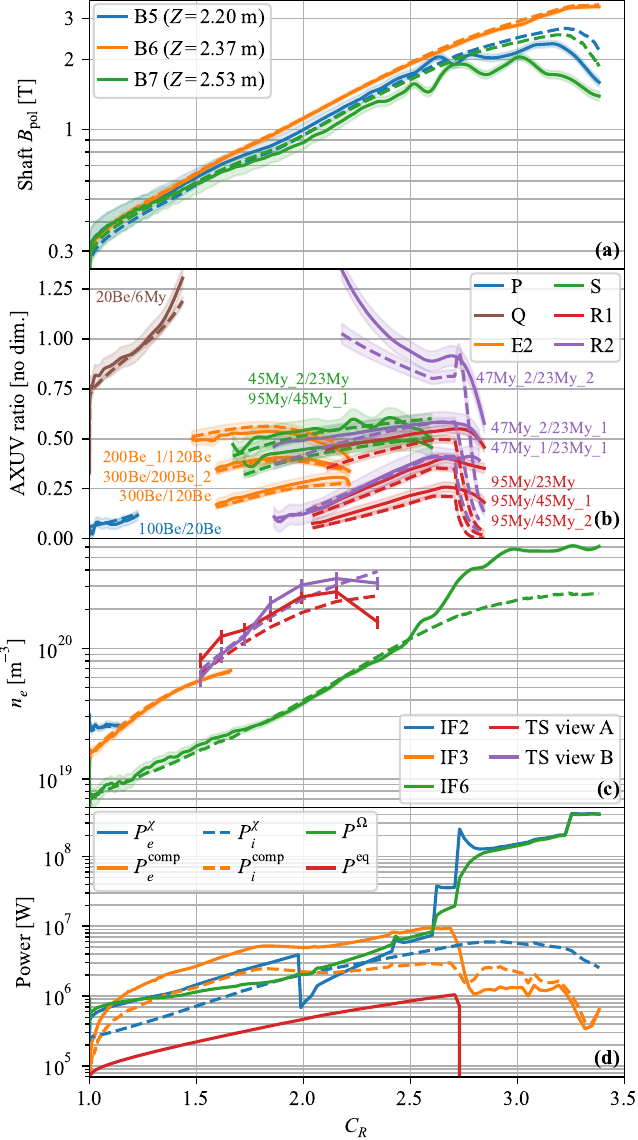} % reverted back to July 30 calibration,
    \caption{(a--c) Comparison of diagnostic data (solid curves) to ISM synthetic results (dashed curves). (a) Poloidal field strength from shaft Mirnov probes B5--7 ($R=0.11$\,m). (b) AXUV filter ratios with diode arrays grouped by color and individual ratios labeled in order of magnitude. (c) Electron density from interferometers (IF) and TS, IF chords are identical to those in Ref.~\cite{Science_of_LM26_Howard_NucFus_2026}. Since TS density was not absolutely calibrated the data were uniformly scaled for each view to best match ISM values. (d) ISM power terms due to thermal transport losses $P^\chi$ (blue), compressional heating $P^\text{comp}$ (orange), Ohmic heating $P^\Omega$ (green), and electron-ion exchange $P^\text{eq}$ (red).
    }
    \label{fig:raw-data-match}
\end{figure}

Figs.~\ref{fig:raw-data-match}a--c show a comparison between diagnostic data and synthetic results from ISM. 
Thermal transport is modeled by effective thermal diffusivities $\hat\chi_e$ and $\hat\chi_i$ which lump together all transport processes. For electrons, a parabolic profile of the form $\hat\chi_e = \hat\chi_{e}^\text{axis} + (\hat\chi_e^\text{LCFS}-\hat\chi_{e}^\text{axis}) \bar\psi^2$ is used, where $\bar\psi$ is the normalized poloidal flux coordinate having value 0 at the magnetic axis and 1 at the LCFS. The ion thermal diffusivity is assumed to be constant and uniform. To achieve a quality fit it is necessary to vary $\hat\chi_e$ in time. For simplicity this is done in a stepwise manner: $\hat\chi_e^\text{LCFS}$ is reduced at 6.5\,ms ($C_R\sim2.0$), followed by increases in both $\hat\chi_{e}^\text{axis}$ and $\hat\chi_e^\text{LCFS}$ at 6.9, 6.95, and 7.15\,ms ($C_R\sim2.6,2.7,3.2$). These step changes are apparent in the electron loss power $P_e^\chi$ shown in Fig.~\ref{fig:raw-data-match}d. We note that a smooth evolution would be able to produce a similarly well-fitting reconstruction. The inferred confinement time at the start of compression is $\tau_{E,e}=6.0$\,ms for electrons and $\tau_{E,i}=5.5$\,ms for ions.

Fig.~\ref{fig:raw-data-match}c exhibits a divergence between the measured and ISM IF6 line-averaged electron density beginning at $C_R=2.5$. Since the ISM model may only lose particles, this implies an ingress of particles to the plasma at that time.

Fig.~\ref{fig:raw-data-match}d shows thermal power balance terms from the ISM reconstruction. Compressional heating becomes the dominant term early in compression at $C_R=1.1$ up until the temperature drops near $C_R=2.7$, after the apparent particle ingress and onset of MHD activity.

The key factors governing the survival of the plasma deep into compression for this experiment are MHD stability, the ingress of impurities, and the diffusion of flux into the liner.  The primarily current-driven MHD stability conditions are dependent on the ratio of shaft current and plasma current, as well as  cross sectional shaping of the plasma determined by the evolving liner shape and plasma current profile. 
The stability of equilibria reconstructed from experimental data are analyzed with the DCON code \cite{glasser-2016-dcon,Brennan2020,Brennan2021}.  
These analyses indicate that controlled ramping of the shaft current during the early-to-mid compression (before the liner completes the coaxial circuit with the shaft) is critical to subsequent stability.  
The ideal ramping scenario would track the inverse inductance $I_{\mathrm{shaft}}(t)\propto L_0/L(t)$ where $L_0$ is the LCFS coaxial inductance at the start of compression.  
The shaft current ramp waveform used on LMC-12 was designed to start below and then rise upwards to reach this ideal curve just before the point where many other LM26 compression shots experienced an MHD event, near $C_R =1.7$~\cite{Science_of_LM26_Howard_NucFus_2026}. 
Reaching the ideal ramp curve when it mattered may have contributed to the stability of the plasma through this mid-compression transition and allowed it to continue to heat efficiently.

Any significant ingress of impurities will impact the survival of the plasma through radiative cooling and increased resistivity. 
For LM26 operations, carefully engineered liner dynamics are a base requirement to minimize impurity injection due to mechanical jetting. Additionally, plasma particle flux to the wall can sputter wall impurities back into the plasma, an effect that will generally increase towards deep compression.    
Flux diffusion into the liner also affects the plasma lifetime directly by consuming flux surfaces, reducing particle and thermal inventory as well as edge currents.

The survival of the plasma may also be impacted by the development of kinetic instability or a change of particle orbits that could increase transport such that it dominates over compressional heating.  Once the energy confinement time falls below the remaining compression timescale the plasma temperature will soon fall.

The overall trend of the LM26 program has been one of pushing the plasma survival deeper into compression, through applying theoretical and empirical considerations to guide operations, with LMC-12 demonstrating the best compression results so far.

\textit{Conclusion---}We have presented evidence that LM26 demonstrated compressional heating to peak temperatures of \mbox{$T_e = 1180 \pm 65\,\mathrm{eV}$}
as measured by AXUV ratio at a radial compression factor of $C_R = 2.75$.
Surpassing fivefold electron temperature increase is validated by TS data (TS view B) showing the same heating trend up to the last measurement of $T_e = 1090 \pm 40\,\mathrm{eV}$ at $C_R = 2.35$, located near the core of the plasma at that moment~\cite{TS_LMC12_Carle_RSI_2026}.  
We infer a meaningful ion temperature increase to $T_i = 459 \pm 33$\,eV based on D-D fusion yield, which the ISM model suggests may be as much as twofold, with a supportive lower bound from C\,V ion Doppler spectroscopy.

A set of improvements were implemented on the LMC-12 campaign that resulted in a peak compressional heating of $P_e^\mathrm{comp}  = 9.6$\,MW, $P_i^\mathrm{comp} = 3.0$\,MW, adding to passively applied internal Ohmic heating of $P^{\Omega} = 8.1$\,MW. 
This improved on the LMC-11 
% result of 
peak heating of $P_e^\mathrm{comp} = 4.3$\,MW, $P_i^\mathrm{comp} = 2.4$\,MW, and $P^{\Omega} = 3.5$\,MW. See End Matter for a more detailed comparison between LMC-11 and LMC-12. 

This significant increase in heating power resulted from only a marginal increase in applied power, demonstrating that sufficient plasma energy confinement and good control of the implosion dynamics and vacuum quality are critical.

The next step in this research program will reconfigure the LM26 machine with a conical converging geometry to achieve even higher compression ratios in a similar collapse time. This will include several planned upgrades to the formation and compression power supplies, as well as doubling the number of formation gas injection valves to provide better control of gas input.  
A new suite of diagnostics will be implemented \cite{NSpecDesign_Carle_RSI_2022} to understand the plasma physics at work within this MTF concept during the plasma evolution toward peak compression. 

The presented observation of simultaneous magnetic flux compression, electron heating, ion heating, and neutron production in a magnetized plasma compressed by a large lithium liner is an important milestone for MTF.  
These results provide confidence in the next phase of the LM26 program as we progress toward 10\,keV.

\textit{Acknowledgments---}The authors thank H. Chohan, J. Stephenson,
S. Bradnam and the UKAEA collaboration
for assistance with MCNP analysis, B. van der Ende from CNL for loan of the $^3\text{He}$ diagnostic, and   
L. Delgado-Aparicio for discussion on soft X-ray diagnostics.

\textit{Data availability---}Diagnostic data will be made available upon reasonable request.

\bibliography{refs}

\appendix

\begin{center}
\bigskip
\small{\textbf{End Matter}}
\medskip
\end{center}

\textit{Details of AXUV analysis---}The AXUV filter ratios shown in Fig.~\ref{fig:raw-data-match}b place the thicker filter of each pair in their respective numerator, resulting in values below unity and a near-monotonic increase with electron temperature in all cases except AXUV~R2 {47My\_2/23My\_2}, which exhibits the opposite behavior in both experiment and ISM. This is caused by the slightly differing views of this particular diode pair in concert with gradients in plasma emissivity.

\textit{Improvements made on LMC-12---}It is important to account for the improvement in compressional heating results for LMC-12 compared to the next-best shot LMC-11~\cite{Science_of_LM26_Howard_NucFus_2026}. Both shots were formed with similar settings and input power levels, resulting in similar values for plasma current and density. LMC-11 showed threefold electron heating from $T_e = 210$\,eV up to 667\,eV, and had a midplane $B_\mathrm{pol}$ increase of 10.8 times. LMC-12 exceeded a fivefold increase in $T_e$ and had a $B_\mathrm{pol}$ increase of 13.4 times.  LMC-12 had slightly higher peak liner velocity (265\,$\mathrm{m/s}$) compared to LMC-11 (255\,$\mathrm{m/s}$), which was helpful but does not fully account for the improvement. The intrinsic improvement of the plasma quality is perhaps best summarized by the ISM estimated initial thermal energy confinement times, which were $\tau_{E,e} = 2.6$\,ms, $\tau_{E,i} = 5.3$\,ms for LMC-11 and increased to $\tau_{E,e} = 6.0$\,ms, $\tau_{E,i} = 5.5$\,ms for LMC-12.  

One likely factor was an improvement in the base vacuum conditions, with LMC-12 having $P_\mathrm{base} =271$\,nTorr just prior to the shot, while LMC-11 was at 350\,nTorr. Residual gas analyzer data showed most of the difference was a significant reduction in both $\mathrm{H}_2$ and $\mathrm{D}_2$, while similar in most other gas components. This could be attributed to an increase in the number of evaporatively deposited lithium coatings during the campaign, with 11 coatings for LMC-12 but only 8 for LMC-11. Additionally, prior to the start of the LMC-12 campaign, the Marshall gun electrode surface was laser cleaned more completely than a previous attempt prior to LMC-11, likely reducing a source of impurities from the wall during the CHI discharge. 

The other change that stands out as important was that LMC-12 began compression later in time than LMC-11 (4\,ms instead of 2.5\,ms). The extra wait time allows the edge density to fall by $\sim$20\%  while the core density only falls by 5\% during this time. A reduction of edge density would be expected to reduce plasma-wall interaction later in compression.
While in principle this extra wait time could allow Ohmic heating to raise the starting temperature of the plasma, this does not align with pre-compression measurements of $T_e$ by AXUV, which show similar starting temperatures of $T_e \approx 200$\,eV at near steady-state.

The one increase in applied power for LMC-12 was that it had 5.4\% higher shaft current at the start of compression, 1.079\,MA instead of 1.024\,MA for LMC-11. This is likely to have reduced the starting value of $\beta_\mathrm{tor}$ proportionally. 
Lastly, while both LMC-11 and LMC-12 had similar delayed ramps in shaft current, which likely helped stabilize the plasma in both cases,  the ramp waveform for LMC-12 was advanced 150\,{\textmu}s earlier than LMC-11, the first pulse fired 1.04\,ms into compression rather than 1.19\,ms, which provided some change in the resistive evolution of the plasma during the phase $1.5 < C_R <2.0$.    

While the analysis so far has yet to pinpoint one dominant factor that best explains the improvement in performance, it remains possible that it was the culmination of several incremental factors.

\begin{figure}[b] % moved float to bottom to avoid broken column balancing
    \centering
    \includegraphics[width=1\linewidth]{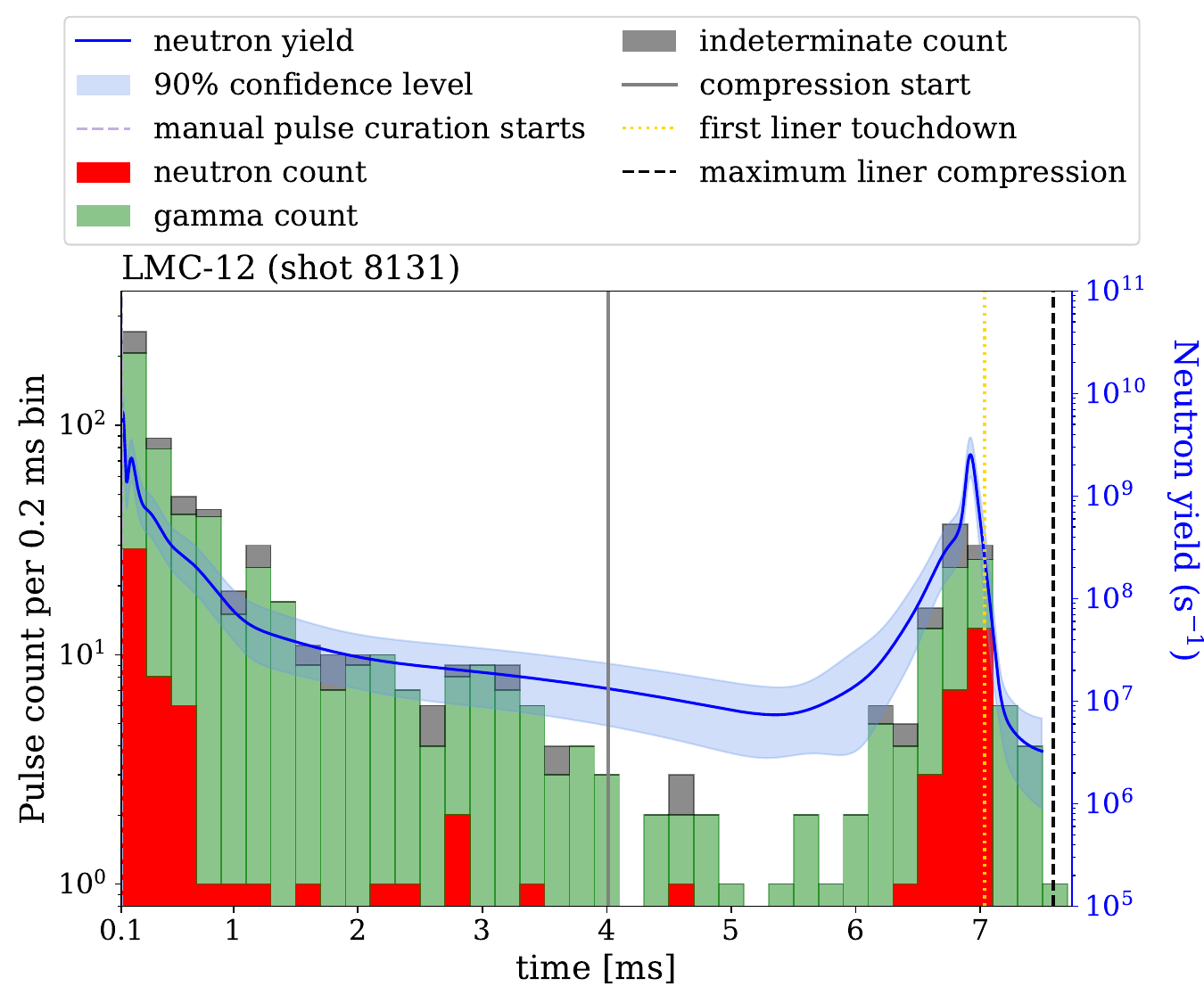}
    \caption{Curated scintillator pulse counts per 0.2\,ms bin (red: neutrons; green: gammas; gray: indeterminate) and neutron yield (blue) with its 90\% confidence band.}
    \label{fig:neutronyield}
\end{figure}

\textit{Neutron yield---}The neutron yield is inferred from liquid scintillator counts. Scintillator pulses are located by peak finding on the filtered, baseline-subtracted waveforms and classified by charge-comparison pulse-shape discrimination (PSD) using a detector-specific cut on the tail-to-total ratio. Pulses followed by another within 70\,ns, or at the digitizer saturation level, receive no PSD identity. Because noise and secondary pulses on the tail can mislead PSD, every pulse from 0.1\,ms after formation was reviewed individually against averaged neutron and gamma templates for its detector and labeled neutron, gamma, indeterminate, or noise; only neutrons are counted (Fig.~\ref{fig:neutronyield}).
The neutron detection times from all scintillators are pooled and the detected rate is reconstructed without fixed time bins using the adaptive kernel estimate of Ref.~\cite{Science_of_LM26_Howard_NucFus_2026}, in which each detection contributes a Gaussian kernel whose width is the interval to its seventh-nearest detection, with a 90\% confidence band set by the local effective number of detections. The yield is this rate divided by the summed detection efficiency, interpolated linearly in time between the eight simulated liner states from the start of compression to peak compression.

When reconstructing the ion temperature from yield and electron density, it is important to have an understanding of the fuel dilution $f_D=n_\mathrm{D}/n_e$. The deuterium density should roughly track the electron density, but $f_D$ has not been directly measured for our plasmas.
A value of $f_D=0.7$ is assumed to be a reasonable representation of the average behavior during early compression, and is consistent with effective ion charge $Z_\mathrm{eff}=1.5$ and average ion charge $Z_\mathrm{avg}=1.2$ used in the ISM.
This will introduce some systematic uncertainty on $T_i$ which is not modeled. The $T_i$ error band is a 90$\%$ confidence interval that comes purely from counting statistics alone and does not incorporate uncertainties from $f_D$ or reconstruction of the density or volume.

\end{document}